\documentstyle[epsf,amstex,onecolumn]{mn.a4}
\loadboldmathitalic 
\title [Isentropic Gas in Galaxy Groups]{Pre-Heated Isentropic Gas in Groups of Galaxies}
\author[M. L. Balogh et al.]
{Mike L. Balogh, Arif Babul, and David R. Patton$^{1}$\\
$^{1}$ Department of Physics \& Astronomy, University of Victoria, Victoria, BC, V8X 4M6, Canada\\
Email: balogh,babul,patton@@uvastro.phys.uvic.ca 
}
\date{\today}
\begin{document} 
\maketitle 
\begin{abstract}
We confirm that the standard assumption of isothermal, shock--heated gas in cluster
potentials is unable to reproduce the observed X-ray luminosity--temperature relation
of groups of galaxies.  As an alternative, we construct a physically motivated model 
for the adiabatic collapse of pre-heated gas into an
isothermal potential that improves upon the original work of Kaiser (1991).  
The luminosity and temperature of the
gas is calculated, assuming 
an appropriate distribution of halo formation times and radiation due to both bremsstrahlung 
and recombination processes.
This model
successfully reproduces the slope and dispersion of the luminosity--temperature relation
of galaxy groups.  
We also present calculations of the temperature and luminosity functions for galaxy
groups under the prescription of this model.  This model makes two strong predictions for haloes with
total masses $M<10^{13}M_\odot$, which are not yet testable with current data: (1)   
the gas mass fraction will increase in direct proportion to the halo mass; (2) the gas temperature
will be larger than the virial temperature of the mass.  The second effect is strong enough
that group masses determined from gas temperatures will be overestimated by about an order of 
magnitude if it is assumed that the gas temperature is the virial temperature. 
The entropy required to match observations can be obtained by heating the gas at the turnaround
time, for example, to about 3 $\times 10^6$ K at z=1, which is too high to be generated
by a normal rate of supernova explosions.
The isentropic model breaks down on the scale of low mass clusters, but this is an acceptable limitation,
as we expect accretion shocks to contribute significantly to the entropy of the gas in such objects.
\end{abstract}

\section{Introduction}\large
The abundance of virialised clusters and groups of galaxies can provide a sensitive probe
of the amplitude of density fluctuations on different scales and, hence, can be used
to determine cosmological parameters.  
This requires, however, a theoretical framework for relating
observable quantities such as the X-ray luminosity (L) and temperature (T) 
of the intra-cluster or intra-group
medium (hereafter referred to as the ICM in both cases, for simplicity) 
to the total mass of the structure.  The most commonly applied model assumes that the gas has been shock
heated to the virial temperature of the isothermal mass.   This results in
a scaling relation $L \propto T^2$ 
\cite{K91,ECF}, which  does not match the observed relation for
clusters, (approximately $L \propto T^3$ \cite{ES91,M98})
or that of groups, which
may be even steeper \cite{P+96}.  It is also true that the $L\propto T^2$ scaling
does not {\em require} isothermal gas; we develop such a model in Appendix \ref{sec-appendix}.

It was originally shown by Kaiser (1991) that, if the gas was 
pre-heated (for example by supernova explosions) 
before falling into the dark matter potential, the observed cluster L--T relation can
be more accurately reproduced.  Assuming adiabatic collapse under these conditions, 
Kaiser's self-similar model   
predicts $L \propto T^{3.5}$.  However, in this case, the radial extent of the gas depends
strongly on the size of the potential; for low mass groups, the gas will extend well beyond
the virial radius and, for high mass clusters, it will be concentrated in the centre.  This
may result in unreasonable gas mass fractions within the virial radius.

Other authors have attempted variations on the self-similar model.  Bower (1994)
explored arbitrary scaling relations of the form $L \propto (1+z)^sM^p$, where s and p
are free parameters, in an attempt to reduce the predicted evolution of the X-ray
luminosity function.  A special case of this treatment is the model of Evrard \& Henry (1991),
which yields the relation $L \propto T^{11/4}$.  Both of these models are based on the
assumption that the core gas possesses a minimum entropy, 
the value of which is determined
by the entropy of the initially pre-heated gas.  Outside this core, it is assumed that the
gas is distributed isothermally.  However, the physical reality of a gas core
has not yet been demonstrated, and the assumption of a largely isothermal profile may 
not be
supported by observations \cite{MFSV}.  Furthermore, as noted by Ponman et al. (1996), this
relation, when extrapolated to lower mass scales, does not fit the X-ray data of galaxy
groups.  

Recently, Cavaliere, Menci \& Tozzi (1997; 1998), have 
constructed a model in which the observed L--T relation on both cluster
and group scales can be reproduced by varying the gas density at the virial radius, according
to the accretion-shock strength, 
as determined by the temperature difference between the infalling
and virialised gases.   In the
present work, we explore the case in which pre-heated gas is assumed to collapse
adiabatically, with the effect of shocks neglected.  This is somewhat similar
to the model of Cavaliere et al.  on group scales, where shocks are expected to be
weak, although we put physical constraints on the total gas mass within virialised
haloes and do not require an isothermal distribution.

The paper is organised as follows.  In \S~\ref{sec-theory}, our cosmological and
structure formation parameters are defined.  The specific models for the gas distributions are 
presented in \S~\ref{sec-models}: in particular 
the isothermal model in \S~\ref{sec-isothermal} and
the isentropic model in \S~\ref{sec-adiabat}.  The  
luminosity--temperature--mass relations and the temperature/luminosity functions are presented
in \S~\ref{sec-results}.  The models are discussed
in terms of their evolutionary predictions, energy requirements and 
failures in \S~\ref{sec-discuss}. The conclusions are summarised in \S~\ref{sec-conc}.

\section{Theoretical Framework}\label{sec-theory}
\subsection{Cosmological Parameters}
We will restrict our analysis in the present work to two simple cosmologies: a standard
cold dark matter (CDM) model with $\Omega_\circ=1$, $h=0.5$, and an open model with
$\Omega_\circ=0.3$, $h=0.75$. The values of $h$ are chosen so that the age of the
universe is equal to about 13 Gyr in both models.  Unless otherwise specified,
the baryon fraction of the universe, $\Omega_b$, will be given by the
big bang nucleosynthesis value $\Omega_b=0.0125 h^{-2}$ \cite{CST}.  

A simple function is fit to the $z=0$ variance of mass fluctuations, $\sigma(M,0)$, from
the CDM power spectrum \cite{BBKS} for the two cosmological models: 
\begin{equation}\label{eqn:fit}
\log(\sigma(M,0)/\sigma_8)=-
\begin{cases}
0.0128+0.298 \log{{M\over M_8}} + 0.0211 (\log{{M\over M_8}})^2&\text{ ; $\Omega_\circ=1$ ,}\\
0.017+0.248 \log{{M\over M_8}} + 0.0187 (\log{{M\over M_8}})^2&\text{ ; $\Omega_\circ=0.3$}\\
\end{cases}
\end{equation}
where the normalisation of the power spectrum 
is determined by the constant $\sigma_8=\sigma(M_8)$, and
$M_8={4 \pi \over 3} \rho_c(0) \Omega_\circ{(8h^{-1}\mbox{Mpc})}^3=5.962 \times 10^{14} h^{-1} \Omega_\circ M_\odot$
is the mass within an $8 h^{-1}$ Mpc sphere, with $\rho_c(0)$ representing the critical density at $z=0$.  
These fits correspond to a spectral index
n ($P(k) \propto k^n$) ranging from $n=-2$ at the low mass end to $n=-1$ for the largest masses
considered in this work.

The linearly extrapolated value of $\sigma(M,0)$ at a redshift $z$ is 
$\sigma(M,z)=\sigma(M,0)g(z)/g(0)$, where the growth term $g(z)$ is given,
to a close approximation in the case of $\Omega_\circ<1$ \cite{VL}, by
\begin {equation}\label{eqn:growth}
g(z)=(1+z)^{-1}
\begin{cases}
1& \text{$\Omega_\circ=1$}, \\
{5 \over 2}\Omega(z) (1+{\Omega(z) \over 2}+\Omega(z)^{4/7})^{-1}& \text{$\Omega_\circ<1$, $\Lambda=0$}\\
\end{cases}
\end{equation}
where $\Omega(z)=\Omega_\circ(1+z)/(1+ \Omega_\circ z)$.

\subsection{Halo Formation Times}\label{sec-LC}
For a halo of a given mass, the density profile of the ICM is related to the shape and 
the depth of the gravitational potential, with the latter depending on the epoch of halo
formation; haloes of a given mass that form at an earlier epoch are denser, and more compact.
This effect has often been neglected in the past, but has been treated recently
by Kitayama \& Suto (1996a, 1996b).
Numerical simulations \cite{NFW2} suggest that the depth of the potential
well, as traced by the circular velocity $V_c$, remains
relatively unchanged after 75 per cent of the cluster mass is in place, since the rest
will be accreted in minor merger events that do not
significantly disrupt the mass already in place.  Thus motivated, we will define the
redshift of halo formation as the redshift at which 75 per cent of the mass has been assembled; the
X-ray temperature of the gas corresponding to a halo of mass $M$ at the epoch of observation
will then be determined by the circular velocity of that halo when its mass was
 equal to $0.75 M$.  We will assume that both gas and dark matter components of merging haloes
mix thoroughly and that the mixed gas immediately settles into hydrostatic equilibrium
after a merger event.

\begin{figure}
\begin{center}
\leavevmode \epsfysize=5cm \epsfbox{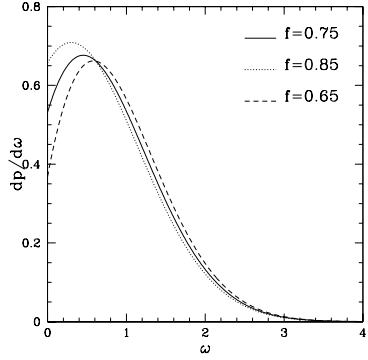}
\end{center}
\caption{The probability distribution of halo formation times in the model of Lacey \& Cole (1993), 
as parameterised by $\omega$, which is related to the halo mass and redshift of formation as described
in the text (\S~\ref{sec-LC}).  The three different curves correspond to three different definitions of
halo formation; the factor $f$ represents the fraction of the final mass (at the observed
epoch) that is virialised at the epoch of formation.}\label{fig-prob}
\end{figure}

\begin{figure}
\begin{center}
\leavevmode \epsfysize=6cm
\epsfbox{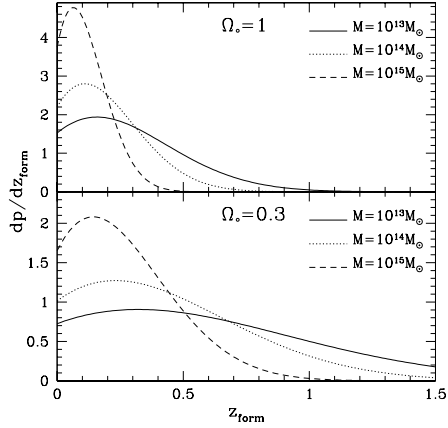}
\end{center}
\caption{The probability distribution of halo formation times as a function
of redshift in the model of Lacey \& Cole (1993), for haloes at $z=0$ with
three different masses, assuming f=0.75.  
The top panel
shows the results for $\Omega_\circ=1$, while the bottom panel shows the results for $\Omega_\circ=0.3$.
Compared to high mass haloes, the low mass haloes have a higher most probable redshift of formation,
as well as a broader distribution of formation times.
\label{fig-pz}}
\end{figure}

An  analytic description of halo formation times, 
applicable to any bottom-up hierarchical model in which structure formation is due to gravitational
instability, was developed by Lacey \& Cole (1993; 1994). 
For a halo of mass M, observed at redshift z, the probability that a  fraction $f$ of 
its mass was virialised at an earlier redshift $z_f$ is
\begin{equation}\label{eqn:prob}
{dp(z_f) \over dz_f}= 
\left[{2\omega(f^{-1}-1)\mbox{erfc}({\omega \over \sqrt{2}})-\sqrt{2 \over \pi}e^{-\omega^2 \over 2}(f^{-1}-2)}\right]{d\omega \over dz_f},
\end{equation}
where the parameter $\omega$ is given by:
\begin{equation}\label{eqn:omega2}
\omega(z_f)={g(0) \over g(z)}{\delta_c (g(z)/g(z_{f})-1) \over \sqrt{\sigma^2(fM(z),0)-\sigma^2(M(z),0)}}.
\end{equation}
Equation \ref{eqn:prob} is only strictly true for the 
special case $\sigma(M,0)=\sigma_8 \left({M \over M_8}\right)^{-1/2}$, 
but the dependence on the shape of the power spectrum
is weak \cite{LC93} and so is neglected for simplicity. 
The dependence on the parameter $f$ is shown in Figure \ref{fig-prob}; the most probable
value of $\omega$ decreases from 0.6 to 0.3 as $f$ increases from 0.65 to 0.85, but this
has a negligible effect on the present calculations.
The distribution of formation times, assuming f=0.75, is shown in Figure \ref{fig-pz} for
three representative mass scales at the present epoch, in both cosmological models.  
Lower mass haloes have a broader distribution of formation epochs, and haloes
in the low density universe model form at earlier times, on average,  than those in the 
critical density model.  This analytic description of halo formation times has been
shown to agree well with numerical simulations \cite{LC94,ENF}.

\subsection{Construction of the Temperature and Luminosity Functions}
The comoving abundance of clusters as a 
function of mass can be determined using the
Press--Schechter (1974) formalism (also Bond et al. 1991), which has
been supported by some numerical simulations \cite{EBW,WEF,LC94,ECF} and is given by
\begin{equation}\label{eqn-mugen}
{dN(>M) \over dM}dM=
\sqrt{2 \over \pi}{\rho_c(0) \Omega_\circ \delta_c \over \sigma(M,z)^2 M}\mbox{exp} \left({-\delta_c^2 \over 2 \sigma(M,z)^2}\right)\left(-{d \sigma(M,z) \over d M}\right)dM,
\end{equation}
where $\delta_c=1.686\Omega(z)^{0.0185}$ is the value of the linearly extrapolated 
critical overdensity 
for spherical gravitational collapse (corresponding to the top-hat filter) 
at redshift z, assuming no cosmological constant \cite{NFW}.   
There is accumulating evidence \cite{BJ,G+98,S+98} that the shape of the mass
function on the scales of groups and clusters, as determined from higher resolution
numerical simulations, 
is not the same as that predicted by the Press--Schechter formalism.  The differences
amount to a factor of 2--3 on both high and low mass scales; nonetheless the Press--Schechter
prediction will be used here, and a study of the error introduced by this procedure
will be deferred.

The differential temperature function is obtained by integrating the mass function,
Equation \ref{eqn-mugen}, over all redshifts:
\begin{equation}\label{eqn:dndt}
{dN(>T) \over dT} = 
\int^{z_{\rm max}}_z {dN(M(T,z_f)) \over dM}{dM(T,z_f) \over dT}{dp(z_f) \over dz_f}dz_f,
\end{equation}
where $z_{\rm max}$ is the redshift at which Comptonization will efficiently 
cool virialised haloes.  The value of $z_{\rm max}=7$ is adopted, but 
 the results are insensitive to this choice for reasonable values (i.e., $z_{\rm max}>5$).  
The cumulative temperature function $N(>T)$ 
is  obtained by numerically integrating Equation \ref{eqn:dndt} over T.  Thus, a relationship
between mass and temperature is required, and this will depend on the thermal history of the ICM.
An analagous procedure is followed to derive the luminosity function, given a relationship
between luminosity and mass.

\section{Models}\label{sec-models}
A singular, truncated isothermal sphere will be assumed for the dark matter potential,
$\rho(r)=\rho_R(r/R_{\rm vir})^{-2}$, where $\rho_R$ is the density
at the virial radius, $R_{\rm vir}$, and is equal to a third of the mean density {\em within}
$R_{\rm vir}$, $\bar{\rho}(R_{\rm vir})$.  This latter quantity is related to the critical 
density at redshift z by $\bar{\rho}(R_{\rm vir})=\Delta_c(z)\rho_c(z)$, and 
$\Delta_c = 78 \Omega(z) + 80 + 300\Omega(z) / (1+15\Omega(z))$ is a fit, 
accurate to better than 2 per cent, to the results of the spherical collapse model as 
presented in Eke et al. (1996).  It will be convenient to define a redshift evolution term,
$F(z)^2=(1+z)^2(1+\Omega_\circ z)\Delta_c(z)/\Delta_c(0)$, so that
\begin{equation}\label{eqn-rhovir}
\rho_R={1 \over 3}\Delta_c(0)\rho_c(0)F(z)^2.
\end{equation}
For $\Omega_\circ=1$, $F(z)^2=(1+z)^3$ and $\Delta_c=178$.

The mass M of a cluster {\em at its formation redshift} $z_f$ will be defined as the mass within
$R_{\rm vir}$.  Thus, the circular velocity $V_c$, which is independent of r, is given
by $V_c^2=GM/R_{\rm vir}$.  The virial temperature is related to $V_c$ by 
\begin{align}
kT_{\rm vir} &= {1 \over 2} \mu m_H V_c^2 \\
         &= 3.1 \left({V_c \over 1000 \mbox{ km s}^{-1}}\right)^2\text{  keV}\notag .
\end{align}
From $M=4\pi\rho_RR_{\rm vir}^3$ the following
relations are obtained:
\begin{equation}\label{eqn-MVc}
\begin{split}
M &=V_c^3 (4\pi \rho_R)^{-1/2}G^{-3/2} \\
  &=2.46 \times 10^{14} h^{-1} F(z_f)^{-1}\left({\Delta_c(0) \over 178}\right)^{-1/2}
  \left({V_c \over 1000 \mbox{ km s}^{-1}}\right)^3 M_\odot 
\end{split}
\end{equation}
and
\begin{equation}\label{eqn-MT1}
\begin{split}
kT_{\rm vir}&=0.5 G\mu m_H (4\pi \rho_R)^{1/3}M^{2/3} \\
        &=1.69 h^{2/3} \left({\Delta_c(0) \over 178}\right)^{1/3} F(z_f)^{2/3} 
\left({M \over 10^{14}M_\odot}\right)^{2/3} \text{  keV .}
\end{split}
\end{equation}

The volume emissivity of the X-ray emitting gas at radius $r$ is given by
\begin{equation}
\epsilon(r)={3 \over 2}{\rho_g \over \mu m_H}{kT_g(r) \over t_{\rm cool}},
\end{equation}
where $T_g(r)$ is the gas temperature, and $t_{\rm cool}$ is its
cooling time.  For the latter, a fit \cite{P96} to
the cooling function of Raymond, Cox \& Smith (1976), constructed
for an H and He plasma with Y = 0.25 
($\mu = 0.59$) and incorporating cooling due to both bremsstrahlung and
recombination, is implemented.  Recombination radiation is important for haloes with
$kT<4$ keV, and its inclusion allows us to consider the mass regime of galaxy groups.  The
expression for the cooling time is
\begin{equation}\label{eqn-tcool}
t_{\rm cool}(r)={C_1 \mu m_H T_g(r)^{1/2}\over \rho_g(r) [1+C_2f_m/T_g(r)]}
\end{equation}
where $C_1=3.88 \times 10^{11}$s K$^{-1/2}$ cm$^{-3}$,  $C_2=5\times10^7$ K and
$f_m$ is a metallicity dependent constant that is 1.0 for solar metallicity 
(adopted here), and 0.03 for zero metallicity.  We only consider haloes in which
not all of the gas within $R_{\rm vir}$ has had time to cool since the halo formed.
The luminosity is given by integrating the volume emissivity out to the virial radius:
\begin{equation}\label{eqn-lum}
\begin{split}
L &=\int_0^{R_{\rm vir}}{4\pi r^2\epsilon}(r)dr \\
  &= {6 \pi k \over C_1(\mu m_H)^2}
\int_0^{R_{\rm vir}}r^2\rho_g(r)^2T_g(r)^{1/2}[1 +C_2 f_m T_g(r)^{-1}]dr. 
\end{split}
\end{equation}

The final density profile of gas that is accreted by a dark matter halo depends on the
density profile of the dark matter itself and also on the
difference between the entropy of the virialised and infalling gas components.  
If the entropy of the infalling gas is low, it will be heated by accretion-shocks
to approximately the virial temperature of the halo,
$T_{\rm vir}$; thus, for our adopted mass profile, the gas distribution will be isothermal.  
Simulations 
show that, indeed, cluster temperature profiles are isothermal to within a factor of
2--3 \cite{NFW,ENF}. 
Alternatively, if the gas is pre-heated so that it has a large initial entropy,
it will be accreted approximately adiabatically, and the gas profile will be
isentropic.  
The purpose of this paper is to consider the properties
of cluster gas under these two extreme scenarios.  For the majority of clusters,
the true physical process is probably intermediate between these models, with the gas becoming
more isothermal and less isentropic with increasing halo mass.

\subsection{Isothermal Model}\label{sec-isothermal}
In this model, we make the common assumption (e.g., Eke et al. 1996) that the gas is distributed isothermally,
with a temperature equal to the virial temperature of the halo.  We include
a distribution of formation redshifts and  recombination radiation, which are often
neglected in studies of this nature. 
The formation redshift $z_f$ of a halo of {\em observed} mass M has been 
defined (\S~\ref{sec-LC}) to be the redshift at which the mass was equal to $fM$.  
Since the gas temperature is assumed not to change between $z_f$ and the epoch of observation,
the observed temperature
and mass are related, from Equation \ref{eqn-MT1}, by
\begin{equation}\label{eqn-MTIso}
kT_g=4.58 \Omega_\circ^{2/3}\left({\Delta_c(0) \over 178}\right)^{1/3}\left({f\over 0.75}\right)^{2/3}
F(z_f)^{2/3}\left({M \over M_8}\right)^{2/3} \mbox{ keV.}
\end{equation}
This gives lower temperatures than the relation of Eke et al. (1996), due to the factor
of f; it is only slightly lower than the normalisation of Pen (1998) which was determined 
from numerical simulations.

If the gas is dissipationless, 
its density profile will match that of the dark matter, i.e., 
$\rho_g(r)=\rho_{\rm g,R}(r/R_{\rm vir})^{-2}$, and $\rho_{\rm g,R}/\rho_R=\Omega_b/\Omega_\circ$.
To avoid the singularity at 
$r=0$ when integrating over the assumed isothermal profile, an arbitrary 
core radius of $r_c = f_c R_{\rm vir}$ is adopted with $f_c=0.1$,
such that $\rho_g(r<r_c)=\rho_g(r_c)$.

Evaluating Equation \ref{eqn-lum} yields the following luminosity--temperature relation:
\begin{equation}\label{eqn:L}
L = 5.83 \times 10^{42}  h {\left ( \Omega_b/ \Omega_\circ \over 0.05 \right )}^2
\left({\Delta_c(0)\over 178}\right)^{1/2}
F(z_f)\left({kT_g \over \mbox{keV}}\right)^2 \left({4/(3f_c)-1 \over 12.33 }\right)
\left [ 1+4.31f_m \left ( {kT_g \over \mbox{keV}} \right ) ^{-1} \right ]\mbox{ ergs s}^{-1}. 
\end{equation}
The luminosity can be expressed as the sum of two components $L_{\rm brem}$ and $L_{\rm rec}$, 
corresponding to the bremsstrahlung and recombination radiation, respectively.  
Substituting $T_g$ 
from Equation \ref{eqn-MTIso},
$L_{\rm brem} \propto M^{4/3}F(z_f)^{7/3}$ and $L_{\rm rec} \propto M^{2/3}F(z_f)^{5/3}$.  
For $\Omega_\circ=1$, this corresponds to $L_{\rm brem} \propto M^{4/3}(1+z_f)^{3.5}$ and
$L_{\rm rec} \propto M^{2/3}(1+z_f)^{2.5}$.

\subsection{Pre-Heated Gas}
\subsubsection{Principles and Definitions}\label{sec-ad_defs}
The equation of state of a polytropic gas is $P=K_\circ\rho_g^{\gamma}$, 
where $P$ is the gas pressure, $\rho_g$ is its density, and
$\gamma$ and $K_\circ$ are constants.  $K_\circ$ is related to the specific 
entropy of the gas, $S$, and
the specific heat capacity at constant volume, $c_v$, by \cite{SS75}
\begin{equation}\label{eqn-entropydefn}
K_\circ={h^2 \over 2 \pi (\mu m_H)^{8/3}}\exp (S/c_v-5/3).
\end{equation}
In the isothermal model 
(\S~\ref{sec-isothermal}), it was
assumed that the gas was shock heated to the virial temperature upon accretion onto 
the dark matter potential; in this case, $\gamma=1$.  At the other extreme, which is the
focus of this section, gas with high
entropy will be accreted adiabatically, and $\gamma=5/3$.  
Evaluation of $K_\circ$ will be discussed further in \S~\ref{sec-Ko}.  

The gas density profile is obtained by solving the equation of hydrostatic equilibrium:
\begin{equation}
{dP \over dr}=-{GM(r) \over r^2}\rho_g(r).
\end{equation}
In principle, $M$ is the mass of both the dark matter and the gas, but we
will neglect the contribution of the gas, the mass of which is only a few percent of the total
(e.g., White et al. 1993).  The resulting
gas density profile within the virial radius is
\begin{align}
\rho_g(r) &= \rho_{\rm g,R}[1+3A^{-1} \ln(R_{\rm vir}/r)]^{1 \over \gamma-1}\mbox{ , }\label{eqn-denprofile}\\
A&={9 \over 2}{c_s^2(R_{\rm vir}) \over V_c^2}={3 \gamma K_o \rho_{\rm g,R}^{\gamma-1} \over (\gamma-1) V_c^2}, \label{eqn-A}
\end{align}
where $c_s(R_{\rm vir})=\sqrt{\gamma K_\circ}\rho_{\rm g,R}^{(\gamma-1)/2}$ is the adiabatic sound speed at the 
virial radius.
This density gradient is significantly shallower than that of isothermal gas 
 and, as a result, the X-ray surface brightness profile is quite
flat near the centre, {\em producing an X-ray core without requiring a core structure
in the underlying dark matter or gas mass distributions}.  Assuming $\gamma=5/3$,
we derive the  ratio of gas mass to dark matter mass
within the virial radius from Equation \ref{eqn-denprofile}:
\begin{equation}\label{eqn-mratios}
{M_g \over M}=0.33{\rho_{\rm g,R} \over \rho_R}A^{-3/2}\mbox{e}^{A}\Gamma(2.5,A),
\end{equation}
where $\Gamma(a,x)=\int^\infty_x e^{-t}t^{a-1}dt$ is the incomplete gamma function.

For an ideal, polytropic gas, the temperature is related to the density by 
\begin{equation}\label{eqn-denT}
\rho_g=\left({kT_g \over \mu m_H K_\circ}\right)^{1 \over \gamma-1}
\end{equation}
so
\begin{equation}\label{eqn-Tprofile}
T_g(r)=T_{g,R}[1+3A^{-1} \ln(R_{\rm vir}/r)]
\end{equation}
where
\begin{equation}\label{eqn-To}
kT_{g,R}={\gamma -1 \over 3\gamma} \mu m_H V_c^2A={2\over 3}{\gamma-1 \over \gamma}AkT_{\rm vir}
\end{equation}
is the gas temperature at the virial radius.
Observationally, the mean cluster temperature measured is an emission weighted temperature, given by
\begin{equation}
T_{\rm em}={\int\epsilon(r)T_g(r)r^2dr \over \int\epsilon(r)r^2dr}.  
\end{equation}
The 
relationship between $T_{\rm em}$ and $T_{g,R}$ is
\begin{equation}\label{eqn-ToTem}
kT_{\rm em}={kT_{g,R}\over A}{\Gamma(5.5,A) \over \Gamma(4.5,A)}\left[{1+{4.31A f_m \over kT_{g,R}}{\Gamma(4.5,A) \over \Gamma(5.5,A)}}\right]
\left[{1+{4.31A f_m \over kT_{g,R}}{\Gamma(3.5,A) \over \Gamma(4.5, A)}}\right]^{-1} .
\end{equation}

The bolometric luminosity is given by substituting the density and temperature
profiles from Equations \ref{eqn-denprofile} and \ref{eqn-Tprofile} 
into Equation \ref{eqn-lum}.  The result is (assuming $\gamma=5/3$):
\begin{equation}\label{eqn-Alum}
L = {6.3 \times 10^{-4} k^{1/2}V_c^{10} \over [C_1\mu m_H G \Delta_c(0) \rho_c(0)]^{1.5} K_o^3 F(z)^3}
\mbox{e}^{A}\Gamma(4.5,A)\left[1 + {7.5 k C_2 f_m \over \mu m_H V_c^2} {\Gamma(3.5,A) \over \Gamma(4.5,A)}\right].
\end{equation}

\subsubsection{Relating $K_\circ$ to the Gas Temperature}\label{sec-Ko}
Here we will assume $\gamma=5/3$.  The entropy constant $K_\circ$, defined in Equation \ref{eqn-entropydefn}, is then related
to the temperature and density of the gas by
\begin{align}\label{eqn-KoTb}
K_\circ&={kT_g \over \mu m_H \rho_g^{2/3}}\\
       &=3.70K_{34}\left({T_g \over 10^6 \mbox{K}}\right)\left({\rho_g \over \Omega_b \rho_c(0)}\right)^{-2/3}\notag
\end{align}
where $K_{\rm 34}=10^{34}$ ergs g$^{-5/3}$ cm$^2$.

The entropy of pre-collapse gas can be increased by heating it to a temperature which will depend
on the gas density at the time of heating.  We will consider heating the gas at the turnaround time,
when the overdense region decouples from the Hubble expansion and begins to collapse; if the
heating takes place instead at some later time during
the collapse of the halo, the gas density will be higher and the amount of heating necessary
will be greater.  At the turnaround time,
the radius of the halo is equal to twice the virial radius e.g. \cite{LC93} and, thus, if we assume the perturbation
to be homogeneous, we have
\begin{equation}
\bar{\rho}_{\rm ta}={1 \over 8}\bar{\rho}(R_{\rm vir})={1 \over 8}\Delta_c(0)\rho_c(0)F(z_f)^2 .
\end{equation}  
Therefore, we can relate the
entropy to the gas temperature at turnaround, $T_{\rm g,ta}$, by
\begin{equation}\label{eqn-Tgta}
K_\circ=314.8K_{34}\left({T_{\rm g,ta} \over 10^6 \mbox{K}}\right)[\Delta_c(0)F(z_f)^2]^{-2/3}.
\end{equation}
In the special case of $\Omega_\circ=1$, Equation \ref{eqn-Tgta} becomes
\begin{equation}\label{eqn-Tgta2}
K_\circ=0.47K_{34}\left({T_{\rm g,ta} \over 10^6 \mbox{K}}\right) (1+z_f)^{-2}\text{ , $\Omega_\circ=1$}.
\end{equation}
The temperature required to obtain a given entropy decreases for lower values of $\Omega_\circ$,
due to the $\Omega_\circ$ dependence of $\Delta_c$.

\subsubsection{The Isentropic Model}\label{sec-adiabat}
Assumptions need to be made about the behaviour of $\rho_{\rm g,R}$ and $K_\circ$ to evaluate
the parameter $A$ in the equations of \S~\ref{sec-ad_defs}.  A simple but physically
unmotivated assumption is the one we made in the isothermal model, 
that $\rho_{\rm g,R} / \rho_R=\Omega_b / \Omega_\circ$.  We discuss interesting results of 
two such models in Appendix \ref{sec-appendix}, but neither produces
an entirely satisfying result.  Furthermore, constraining $\rho_{\rm g,R}$ at
$r=\infty$ is not a satisfactory constraint because gas cannot be assumed to be in
hydrostatic equilibrium beyond a distance approximately equivalent to $c_s/H_\circ$,
where $H_\circ$ is the Hubble constant; this distance is equal to a cluster virial
radius within a factor of order unity.  Beyond this radius, gas will be infalling and
its distribution is not represented by our model.

Instead, we will assume that the entropy constant $K_\circ$ is independent of
halo mass, and that the total mass of gas in the cluster is determined by the
amount that could have been accreted at the
adiabatic Bondi accretion rate \cite{Bondi}, assuming that the accreted gas has
already been pre-heated, and accretes isentropically:
\begin{align}\label{eqn-ABR}
\dot{M}&= 4\pi\lambda G^2M^2(\gamma K_\circ)^{-3/2}\rho_g^{{3 \over 2}(5/3-\gamma)}, \\
       &\approx1.86\pi\lambda G^2M^2K_\circ^{-3/2}, \text{ $\gamma=5/3$}\notag
\end{align}
where $\lambda=0.25$
is the dimensionless accretion rate and $\rho_g$ and $T_g$ are the density and 
temperature of the gas being accreted.   {\em We assume $\gamma=5/3$ for the remainder
of the derivation}.  If the simplifying assumption is
made that the accretion rate has remained roughly constant, 
the amount of gas mass accreted by the {\em observed} redshift z, $M_g(z)$, can be determined from
\begin{equation}\label{eqn-bondi}
{M_g(z) \over M}=0.040 {M \over 10^{14}M_{\rm \sun}}\left({K_\circ \over K_{\rm 34}}\right)^{-3/2} {t(z) \over 10^{9}\mbox{ years}},
\end{equation} 
where $t(z)$ is the Hubble time at redshift z.  Thus, we predict a strong mass dependence of 
the gas mass fraction, $M_g / M \propto M$, for the lowest mass haloes (see below for the mass limit).
There is currently insufficient data available to accurately test this prediction, but there are indications
that the gas mass fraction increases significantly from elliptical galaxies to groups to clusters \cite{DJF}.

The model gas mass fraction, from Equation \ref{eqn-bondi}, will exceed $\Omega_b / \Omega_\circ$ for clusters
with circular velocities greater than the critical value $V_{\rm crit}$:
\begin{equation}\label{eqn-Vcrit}
V_{\rm crit}=998h^{1/3} \left({\Delta_c(0)\over 178}\right)^{1/6}\left({\Omega_b / \Omega_\circ \over 0.05}\right)^{1/3} 
\left({K_\circ \over K_{\rm 34}}\right)^{1/2} \left({t(z) \over 10^{9} \mbox{years}}\right)^{-1/3}F(z_f)^{1/3}\mbox{ km s}^{-1}.
\end{equation}
For $\Omega=1$, this corresponds to a halo mass of
$M_{\rm crit}=3.78 \times 10^{14} h\Omega_b (K_\circ/K_{34})^{3/2}(1+z_f)^{3/2} M_\odot$.  
Above this mass limit, the Bondi accretion rate will be ignored, and the condition
$M_{\rm g} /M = \Omega_b/\Omega_\circ$ will be imposed.  
This is in reasonable agreement with observations (e.g. Edge \& Stewart 1991; Evrard 1998) which show that
gas mass fractions are roughly constant among clusters, though there is a considerable amount of
scatter.  The critical halo mass should be compared
with the Jeans mass at the turnaround time, which is defined by
\begin{equation}
M_J={4 \pi \over 3}\Omega_\circ \rho_c(z_{\rm ta}) \lambda_J(z_{\rm ta})^3
\end{equation}
where the Jeans length is $\lambda_J=t(z) c_s$.  Using the relations $t_{f}=2t_{\rm ta}$ and $4(1+z_f)^3=(1+z_{\rm ta})^3$, (the
latter valid only for $\Omega_\circ=1$),
the Jeans mass of a halo at turnaround can be expressed in terms of its formation redshift as
$M_J=2.4 \times 10^{14} h \Omega_b(K_\circ/K_{34})^{3/2}(1+z_f)^{3/2} M_\odot$.  Thus, we obtain
$M_{\rm crit}/M_J=1.6$ which is independent of of $z_f$, $K_\circ$ and $h$.  Since the Jeans mass, which
determines the scale at which pressure effects are important relative to gravitational collapse, is not
much smaller than $M_b$, our assumption of adiabatic Bondi accretion is not unreasonable.

For low mass haloes, with circular velocities less than the critical one given
in Equation \ref{eqn-Vcrit}, the constant $\rho_{\rm g,R}$ is defined parametrically
(through $A$) by
\begin{equation}\label{eqn-V<Vcrit}
\mbox{e}^{A}\Gamma(2.5, A)=6.85h\left({\Delta_c(0) \over 178}\right)^{1/2}F(z_f){t(z) \over 10^{9}\mbox{years}}.
\end{equation}
Thus, the parameter $A$ is independent of $V_c$, but does depend on the redshift of formation in
a non-trivial manner.  From Equation \ref{eqn-A}, this requires $\rho_{\rm g,R} \propto V_c^3$.
Since $A$ is independent of $V_c$ then, from Equations \ref{eqn-MVc}, \ref{eqn-To} and \ref{eqn-ToTem} 
the isothermal M--T relation $T_{\rm em} \propto M^{2/3}$ is again obtained, but the constant of
proportionality changes and the dependence
on formation redshift is more complex.  Allowing for the fact that, on average, higher mass
haloes form at lower redshifts than low mass haloes, the M--T relation will be very slightly
modified.  The luminosity dependence on mass, however, 
being more sensitive to the gas distribution, differs significantly from the isothermal
case.  Considering bremsstrahlung radiation only, for example, 
$L \propto V_c^{10}F(z)^{-3} \propto T_{\rm em}^5 F(z)^{-3}$.  Thus, the important result
is obtained that, by assuming adiabatic collapse of gas and a gas mass fraction
which is proportional to the total mass, $L \propto T_{\rm em}^5$.   

For high mass haloes, with circular velocities greater than the critical value, the situation
is considerably more complex.  To solve $\rho_{\rm g,R}$, the parameter A must be determined
from the following equation:
\begin{equation}\label{eqn-V>Vcrit}
\mbox{e}^{A}\Gamma(2.5,A)=(7.5K_\circ)^{1.5}\left({\Omega_b \over \Omega_\circ}\right)[\Delta_c(0)\rho_c(0)]V_c^{-3}F(z)^2.
\end{equation}
Thus, A (and, hence, $\rho_{\rm g,R}$) depend on both the mass and formation redshift in 
a complex manner.  Furthermore, Equation \ref{eqn-V>Vcrit} is not soluble for all $V_c$.  The
left hand side of the equation approaches 1.33 as A approaches zero, which means that
solutions can only be found for haloes with circular velocities that satisfy
\begin{equation}\label{eqn-Vmax}
V_c<1366h^{2/3} \left({K_\circ \over K_{\rm 34}}\right)^{1/2}\left({\Omega_b /\Omega_\circ \over 0.05}\right)^{1/3}
\left({\Delta_c(0) \over 178}\right)^{1/3}F(z)^{2/3} \mbox{ km s}^{-1}.
\end{equation}
The model breaks down because the gas becomes increasingly concentrated toward the cluster
centre as mass increases and, eventually, the density at the virial radius becomes zero.
Since shock heating is expected to dominate the gas entropy for large masses, this is
an acceptable limitation.  Alternatively, a solution exists for any mass if the entropy
constant $K_\circ$ is made sufficiently high. 

The bolometric luminosity, L, is then determined from Equation \ref{eqn-Alum} after solving
for A from Equation \ref{eqn-V<Vcrit} or \ref{eqn-V>Vcrit}.
Similarly, the emission weighted temperature is computed from equations \ref{eqn-To} and \ref{eqn-ToTem}.  Note that 
$L$ and $T$ do not scale simply with $h$, due to the complex $h$ dependence of $A$.

The restriction on circular velocity presented in Equation \ref{eqn-Vmax} can be expressed
in terms of temperature and luminosity limits as:
\begin{equation}\label{eqn-Tcrit}
kT_{\rm crit}<6.9h^{4/3} {K_\circ \over K_{\rm 34}}\left({\Omega_b /\Omega_\circ \over 0.05}\right)^{2/3}
\left({\Delta_c(0) \over 178}\right)^{2/3}F(z)^{4/3} \mbox{  keV}
\end{equation}
and
\begin{equation}\label{eqn-Lcrit}
\begin{split}
L_{\rm crit}  <1.61 \times &10^{44}h^{11/3} \left({K_\circ \over K_{\rm 34}}\right)^2 F(z)^{11/3}
\left({\Omega_b /\Omega_\circ \over 0.05}\right)^{10/3}\left({\Delta_c(0) \over 178}\right)^{11/6} \times \\
&\biggl[1+0.79h^{-4/3}f_m 
\left({K_\circ \over K_{\rm 34}}\right)^{-1}\left({\Omega_b /\Omega_\circ \over 0.05}{\Delta_c(0) \over 178}\right)^{-2/3}F(z)^{-4/3} \biggr]\mbox{ ergs s}^{-1}. 
\end{split}
\end{equation}

\section{Results}\label{sec-results}
\subsection{The Luminosity--Temperature Relation}\label{sec-LT}
\begin{figure*}
\begin{center}
\leavevmode \epsfysize=12cm \epsfbox{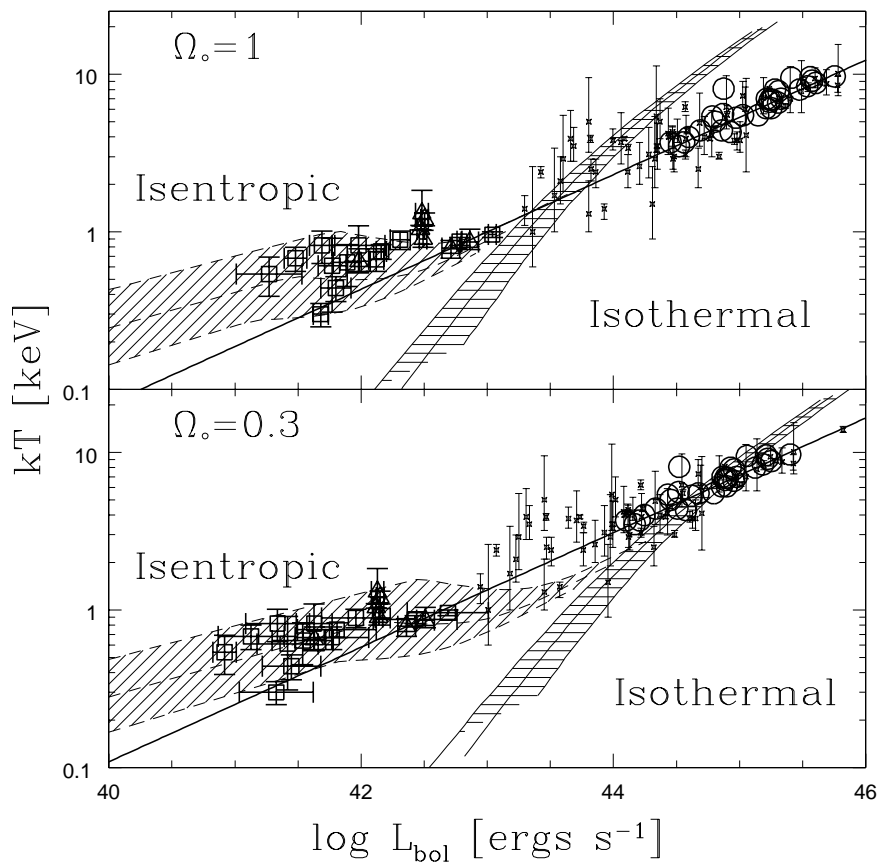}
\end{center}
\caption{The luminosity--temperature relation, with data from David et al. (1993,1995 stars),
Markevitch (1998, circles), Ponman et al. (1996, open squares, resolved sources
only) and Mulchaey \& Zabludoff 
(1998, open triangles, Raymond--Smith model determinations with half-solar metallicities).
The error bars are omitted from the Markevitch data, as they are smaller than the data
symbols.  
The isothermal model and
isentropic model are plotted; the hatched region reflects the range of formation
redshifts for clusters of a given mass.  Only the isentropic model matches the slope and dispersion of 
the group data well; the normalisation of this model depends on the initial entropy, which we determine
by fitting to the group data.  The straight, solid line represents the  model of
Evrard \& Henry (1991).  Although this matches the cluster data well, 
it underpredicts the temperature of low luminosity groups.
The models and data do not scale with $h$ in an identical manner, as discussed in the text.
The difference between the top panel ($\Omega_\circ=1, h=0.5$) and bottom panel ($\Omega_\circ=0.3, h=0.75$)
is due primarily to the different values of $h$ used, and not $\Omega_\circ$.
\label{fig-LT}}
\end{figure*}

The L--T relation\footnote{The emission weighted temperature of the model is always considered
when comparing with observations.} for both models and cosmologies is shown in Figure \ref{fig-LT}; temperature
is plotted on the vertical axis, somewhat contrary to convention, since this quantity
has the dominant observational uncertainty.
For each model, three lines are drawn, representing the range of formation
redshifts between  $z_f=0$ and the
$1\sigma$ upper limit (i.e., where the probability that the cluster formed at a lower 
redshift is equal to 68.3 per cent).  The central line represents the calculation for the 
most probable formation redshift.  The data plotted are those of David et al. (1993;1995, stars), 
Markevitch (1998, circles),
Ponman et al. (1996, open squares)\footnote{Only fully resolved observations are considered 
(i.e., with a quality index of 1).} and 
Mulchaey \& Zabludoff (1998, open triangles)\footnote{We use the temperatures determined using the 
Raymond--Smith model with the metallicity fixed at half solar for all groups except NGC5846, for which
this temperature is unconstrained.  In this case, we adopt the low metallicity determination.}.   All data points
are unique; where there is overlap we choose the Ponman et al. group data over the Mulchaey \& Zabludoff
data, and the Markevitch cluster data over that of David et al.
 
It is crucial to note that the data and models do not scale with $h$ in the same way;
for this reason, $h$ has not been left as a free parameter.  The observed luminosities
scale as $L \propto h^{-2}$, while the isothermal model\footnote{Recall, from \S~\ref{sec-adiabat}, that the values of $L$ and $T$ in the isentropic
model do not scale with $h$ in a simple manner.} 
scales as $L \propto h\Omega_b^2$, from
Equation \ref{eqn:L}.  Thus, if $\Omega_b$ is fixed, $L\propto h$, while
if the big bang nucleosynthesis value is adopted, as we do here, $L\propto h^{-3}$. 
In either case, the ratio of model to observed luminosity is $h$ dependent.
This accounts for most of the difference
between the matching of the model and the data in the 
$\Omega_\circ=1$ cosmology (top panel, $h=0.5$) and the 
$\Omega_\circ=0.3$ cosmology (bottom panel, $h=0.75$). 

The slope of the isothermal model is too steep to match the 
cluster data, even for haloes with temperatures greater than a few keV.  
Due to the effects of recombination radiation, 
it steepens even further for $kT<4$ keV,
approaching $L \propto T$ and, thus, completely fails to match the low-luminosity, group
data.  Furthermore, for luminosities less than about $10^{43}$ ergs s$^{-1}$ the gas will have had time
to cool efficiently since the formation of the halo and will no longer be emitting 
X-ray radiation.

For the isentropic model, the entropy constant, $K_\circ$, has been determined by matching 
the L--T relation to the available group data of Ponman et al. (1996) and Mulchaey \& Zabludoff (1998).    
This constraint requires a value of $K_\circ=(0.37 \pm 0.1) K_{\rm 34}$ 
which, from Equation \ref{eqn-Tgta2}, can be obtained by heating the gas at the turnaround
time to a temperature of about $T=7.9 \times 10^5 (1+z_f)^2$ K, for $\Omega_\circ=1$.
For $z_f\approx1$, the required gas temperature is $3.1 \times 10^6$ K; this 
is reduced to $T=1.8 \times 10^6$ K for $\Omega_\circ=0.3$.  

The slope of this model for the lowest masses ($L \propto T^5$) is certainly 
consistent with most of the group data.  Ponman et al. (1996)
claim the slope of their data is best fit by $L \propto T^{8.2}$, but there is a great
deal of scatter and there are significant outliers.  Including the data of Mulchaey
\& Zabludoff (1998) we obtain a best fit relation of $L \propto T^{3.5}$, but the
chi-squared value indicates a poor fit, so the errors on the slope are not well determined.  
Clearly, more data with smaller errors are required
to constrain the slope in this region.
It must also be noted that the Ponman et al. (1996) sample is a collection of Hickson
compact groups, which are atypical groups.  It is,
in fact, remarkable that they lie in the same area of the L--T plane as the Mulchaey \&
Zabludoff (1998) group data.
	
The change in model slope
at $L \approx 10^{42}$ ergs s$^{-1}$ is due to the change in gas mass constraint,
from the amount of mass accreted at the adiabatic Bondi rate to a fixed value 
(at $V_c=V_{\rm crit}$; see \S~\ref{sec-adiabat}).  As the halo mass is increased beyond this
value, the central gas density increases rapidly, and the gas density at the
virial radius drops.  Since the density and temperature
gradients are initially fairly shallow, the total emission weighted
temperature is dominated by the greater volume of gas at large radii; the lower
temperature of this gas
approximately compensates for the central increase in temperature, and the total emission-
weighted temperature remains roughly constant.   The luminosity, on the other hand, 
is more strongly dependent on 
density  and, thus, continues to increase with $V_c$ (although not as strongly as for
haloes with $V_c<V_{\rm crit}$).  
In this mass regime, the model does not fare as well, generally
underpredicting the observed temperatures; this is not unexpected, as accretion shocks will
begin to become important for haloes of this size.  One possible model for the
structure in these haloes is that the pre-heated gas dominates in the
central regions, while accretion-heated gas dominates elsewhere.
A simple model of this type is described by Evrard \& Henry (1991), and is represented by
$\log kT=4/11 \log(h^2L)-15.42$ when normalised to the cluster data.
This model is shown here as the straight, solid line, which our isentropic model approaches 
asymptotically.  The Evrard \& Henry model matches the slope of the cluster
data quite well, but
underpredicts the temperature of most groups and the lowest mass clusters.  

It has been shown \cite{F+94,AE98,M98} that the 
dispersion in the L--T relation can be reduced to 
a very small value (an r.m.s. dispersion of about 0.11 in log L at a given T) by excluding
cooling flow regions from the observed X-ray data.  
The predicted ($1\sigma$) range in log L for a given T for high mass clusters under the
assumptions of the isothermal model, due to the distribution of halo formation times, 
is 0.07-0.14, 
consistent with this data (though, of course, the slope is wrong).  
The isentropic model predicts a larger range of about 0.3 for low luminosity 
clusters and groups, and this
range encompasses most of the group data shown in Figure \ref{fig-LT}.  This suggests that
the observed dispersion in the L--T relation may be primarily due to the distribution of halo
formation times.

\subsection{The Mass-Temperature Relation}
\begin{figure}
\begin{center}\leavevmode\epsfysize=8cm
\epsfbox{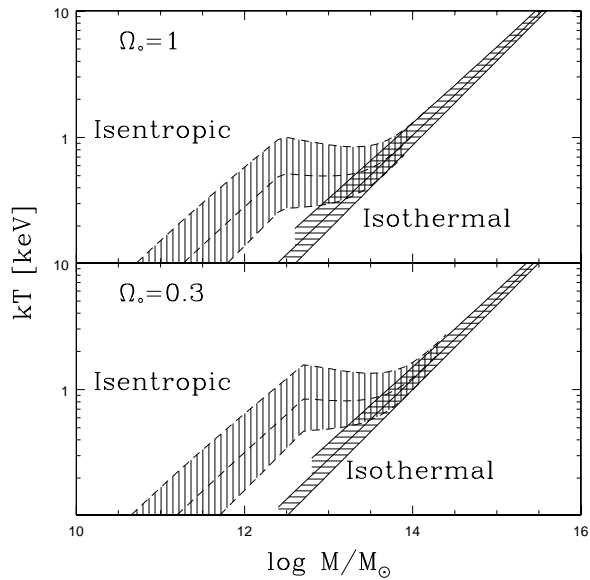}
\end{center}
\caption{The mass--temperature relations for $\Omega_\circ=1$ (top) and $\Omega_\circ=0.3$ (bottom).
The two models shown in each panel are the isothermal and isentropic models
discussed in the text.  The hatched region reflects the range of formation redshifts
for clusters of a given mass.  On the scale of groups of galaxies, both models predict
$M \propto T^{1.5}$, but the mass corresponding to a given temperature is an order of
magnitude smaller in the isentropic model, compared with the isothermal model.
\label{fig-MT}}
\end{figure}

The mass derived from the X-ray temperature of group-scale structures is strongly 
dependent on the thermal history of the gas, as shown in Figure \ref{fig-MT},
where the mass--temperature relation is plotted for the isothermal
and isentropic models; as in Figure \ref{fig-LT}, the hatched region represents the dispersion in the relation
due to the distribution of halo formation times. 
At low masses, $M< 10^{13}M_\odot$, the gas mass corresponding to a given temperature
is about an order of magnitude smaller in the isentropic model than in the isothermal model:
the mass of haloes with $kT<0.4$ keV will be greatly overestimated by assuming
the gas has been accretion-shocked to the virial temperature.  Furthermore, the dispersion
of the M--T relation is significantly larger in the isentropic model; thus it will be 
difficult to determine the mass of individual, low temperature structures to better than a factor
of ten or so from the X-ray observations alone.  In the isothermal case, for high
temperature clusters, the relation is much tighter, and the mass corresponding to a given
temperature can be determined to within a factor of two.

\begin{figure}
\begin{center}\leavevmode\epsfysize=8cm
\epsfbox{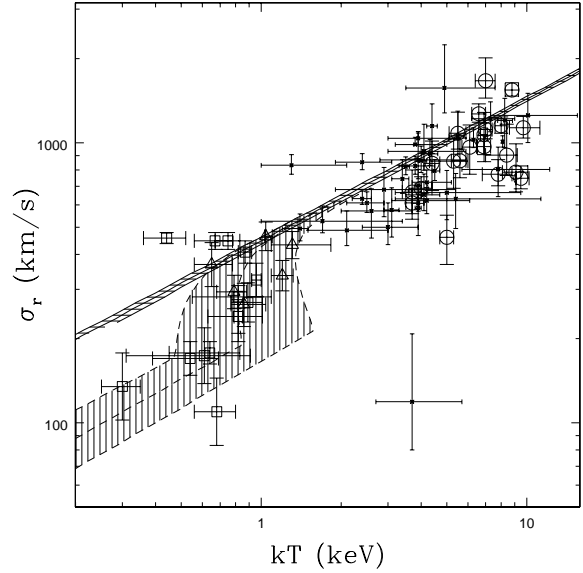}
\end{center}
\caption{We show the relation between line--of--sight velocity dispersion and gas temperature
for the isentropic and isothermal models, with $\Omega_\circ=0.3$.  The width of the hatched
region reflects the range of formation redshifts; the model with the wider band is the
isentropic model.  The data sources for the temperatures are the same
as for Figure \ref{fig-LT}; sources of the velocity dispersions are described in the text.
\label{fig-Tsig}}
\end{figure}

The most readily observable, independent parameter related to the halo mass is the line of sight
velocity dispersion, $\sigma_r$.    Assuming isotropic orbits, we use the relation $V_c=\sqrt{2}\sigma_r$ to
compare our model predictions with observations.  We show, in Figure \ref{fig-Tsig}, the relation
between $\sigma_r$ and gas temperature predicted by the isothermal and isentropic models for $\Omega_\circ=0.3$.  
The group data 
plotted are the same as in Figure \ref{fig-LT}; velocity dispersions for the Mulchaey \& Zabludoff (1998)
are taken from Zabludoff \& Mulchaey (1998).  For the cluster data, 
velocity dispersions (not available for all of the data) are taken from Fadda et al. (1996) and various
sources within the compilation of White et al. (1997).  In the rare cases where uncertainties are not
quoted, we assume 20\% errors.

The lowest temperature group data generally lie well
below the isothermal model prediction; however, there is a great deal of scatter, and 
insufficient data below $kT=0.4$ keV to
provide a convincing test of our isentropic model prediction.  Furthermore, even the massive cluster
data tend to lie toward higher temperatures at a given $\sigma_r$, which may indicate that the galaxy orbits
are not exactly isotropic.  The most discrepant point
is the cluster AWM4, which seems to have an unusually low velocity dispersion of 119 km/s (Fadda et al. 1996)
for its temperature of 3.7 keV. 
Nonetheless, the data suggest a strong drop in $\sigma_r$, relative to the isothermal model, for haloes
with temperatures below about
0.8 keV, in accord with our isentropic model.  The notable exception to this trend is 
HCG 15, which has kT=0.44 keV and $\sigma_r$=457 km s$^{-1}$ (Ponman et al. (1996).

The difference between the isothermal and isentropic models in Figures \ref{fig-MT} and \ref{fig-Tsig}
is somewhat sensitive to the mass accretion rate $\lambda$ we have assumed in Equation \ref{eqn-ABR}.  A lower
accretion rate (i.e. $\lambda<0.25$) will preserve the slope of the M--T relation, while altering
the normalisation, as the total gas mass within a halo of a given mass will be lower than we have assumed
here.  In this case, the temperature will drop.  Our choice of $\lambda=0.25$ is the critical value determined
by Bondi (1951) for the case of spherical, adiabatic accretion by a point source; we are currently exploring
a more realistic accretion model, and indications are that the proper $\lambda$ may be somewhat lower than we
have used here.

\subsection{The Temperature Function}
\begin{figure}
\begin{center}\leavevmode
\epsfysize=8cm
\epsfbox{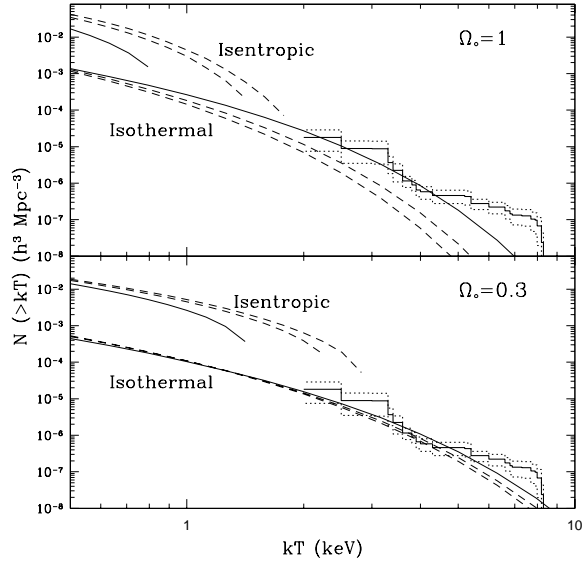}
\end{center}
\caption{The cumulative temperature function for $\Omega_\circ=1$ (top) and $\Omega_\circ=0.3$ (bottom).
The isothermal and isentropic models are shown at three redshifts,
z=0 (solid line), z=0.3 z=0.5 (dashed lines).  
The z=0 models are normalised to the data (Henry \& Arnaud 1991, with errors 
from Pen 1998) at kT=3.4 keV. For the isothermal case, evolution at the high temperature end 
is strongly negative for $\Omega_\circ=1$ and only weakly so for $\Omega_\circ=0.3$.  The isentropic model predicts
some positive evolution for objects with $kT<2$ keV but, as noted in
the text (\S~ \ref{sec-adiabat}), it breaks down for larger mass haloes.
\label{fig-N>T}}
\end{figure}

The cumulative temperature functions for the two models are shown in Figure \ref{fig-N>T}.  
The $\Omega_\circ=1$ cosmology is shown in the top panel, and the $\Omega_\circ=0.3$ results in
the bottom panel.  The $z=0$ model is shown as the solid line, and the dashed lines
represent the results at redshifts of 0.3 and 0.5.  
The cluster data from Henry \& Arnaud (1991), as analysed 
by Eke et al. (1996) and with 1$\sigma$ 
errors taken from Pen (1998), is plotted as the step function in this figure.   
This data consists of the 25 X-ray brightest clusters in the {\em HEAO-1 A2} survey, 
as detected in the 2--10 keV band.

The {\em r.m.s.} mass fluctuation in spheres of size $8 h^{-1}$ Mpc, $\sigma_8$, is 
determined by requiring the cumulative temperature
function at z=0 to match the observations at $kT=3.4$ keV, where the errors are smallest.  
For the isothermal model, $\sigma_8=0.802$ for $\Omega_\circ=0.3$, and $\sigma_8=0.492$
for $\Omega_\circ=1$.  These values are lower than those found by Eke et al. (1996),
primarily as a result of including the distribution of formation times.  Neglecting this
(i.e., assuming each cluster formed at the redshift at which it is observed)
overestimates $\sigma_8$ by 10--20 per cent, depending on cosmology (a larger effect for $\Omega_\circ=0.3$).
The effects of using a slightly different point of normalisation and 
mass--temperature relation are much weaker, affecting the results by about 1 per cent.
  
The isentropic model is shown in Figure \ref{fig-N>T} for the same three
redshifts.  For this model, $\sigma_8$ cannot be calculated in the same way, as the model 
breaks down for temperatures greater than 1--2 keV, so the values determined above for
the isothermal model are used.  In this case, evolution is mildly {\em positive}; i.e., there are
more objects of a given temperature at z=0.5 than at z=0, by a small amount that is
roughly independent of cosmology.  Relative to the isothermal temperature function,
the isentropic model predicts about ten times more objects with temperatures less than 1 keV.
This is also evident from Figure \ref{fig-MT}: the masses of haloes with a given X-ray
temperature are much smaller in the isentropic model, and less massive objects are
more numerous.

\subsection{The Luminosity Function}
\begin{figure}
\begin{center}\leavevmode
\epsfysize=8cm
\epsfbox{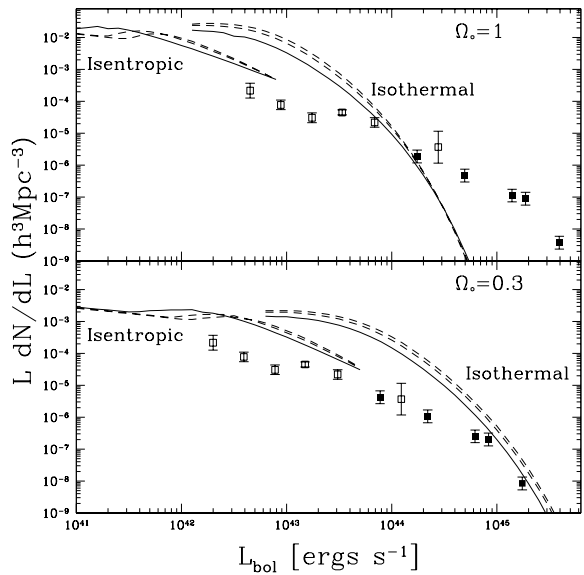}
\end{center}
\caption{The differential luminosity functions for $\Omega_\circ=1$ (top) and $\Omega_\circ=0.3$ (bottom).
The two models shown in each panel are the isothermal and isentropic models
discussed in the text, at z=0 (solid line), z=0.3 and z=0.5 (dashed lines).  Very little evolution is exhibited
in either model or cosmology.  The flattening of the isothermal luminosity function at low $L$
results from efficient cooling of the gas in low mass haloes.
The data are from Rosati et al. (1998, open squares) and Henry \& Arnaud (1991, filled squares).  
The normalisation of the models relative to the data is sensitive primarily to the value of $h$, since
the two do not scale the same way.
\label{fig-Ldif}}
\end{figure}

Although temperature data do not yet exist for a complete sample in the 
low mass regime, a differential luminosity function is available from the
{\em ROSAT} Deep Cluster Survey (RDCS, Rosati et al. 1998) to luminosities as low as $2 \times 10^{42}$
ergs s$^{-1}$.  We correct these 0.5--2 keV band luminosities to bolometric luminosities by
dividing them by the factor $\exp(-0.5/kT) - \exp(-2/kT)$, where the temperature is estimated
from the Evrard \& Henry (1991) relation shown in Figure \ref{fig-LT}. 
The lowest redshift data from the RDCS ($0.045<z<0.25$) are shown as the
open squares in Figure \ref{fig-Ldif}; the solid squares plotted on this figure
are from Henry \& Arnaud (1991).  The isothermal and isentropic models
are shown for both cosmologies.  For each model, the solid line is the $z=0$ model, while the
dashed lines represent the results at $z=0.3$ and $z=0.5$; very little evolution is exhibited in either cosmology by
either model.  This is consistent with the RDCS survey results, which show no evolution of the
luminosity function out to z=0.8.

The slope of the luminosity function corresponding to the isothermal model is too
steep to match the data in the high luminosity regime, especially for the $\Omega_\circ=1, h=0.5$ cosmology.  
As discussed earlier, the normalisation difference between the models and the data is
mostly sensitive to $h$.
The flattening of this slope at low luminosities is due to the fact that we do not include
haloes in which the gas has had sufficient time to cool since the formation of the halo;
efficient cooling in low mass objects restricts luminosities to those greater
than about $10^{43}$ ergs s$^{-1}$.

The isentropic model, on the other hand, reproduces the flat
slope ($d\log(dN/dL) / d\log(L)=-1.88$) of the RDCS data well, though 
the normalisation is too high. 
However, the major success of the L--T relation predicted by this model occurs at
luminosities lower than those probed by the RDCS ($L<10^{42}$ ergs s$^{-1}$)
 where the gas mass fraction is proportional to the total mass (\S~\ref{sec-adiabat}).  
We predict that the slope of the luminosity
function in this mass range will flatten to $d\log(dN/dL) / d\log(L)=-1.33$.

\section{Discussion}\label{sec-discuss}
Recently, Henry (1997) published a detection of mild negative evolution of
the cluster temperature function out to z=0.3, particularly at the hot end, $kT>5$ keV.
This is marginally inconsistent with the isothermal model in both cosmologies in Figure \ref{fig-N>T}, 
as the $\Omega_\circ=0.3$ 
temperature function shows too little evolution, while the evolution
predicted by the $\Omega_\circ=1$ model is too strong.  For this reason, Henry (1997) claims
$\Omega_\circ=0.5$ fits the data best, but there are few points and the error bars are 
large.  The isentropic model predicts very little evolution in both
cosmological models, but it is only valid at low temperatures for which there is currently
no data.  Since the redshift evolution in this case has such a weak cosmological dependence,
estimations of cosmological parameters from X-ray abundances will best be made
from the number evolution of the most massive clusters, and not from the abundance of
lower mass clusters and groups.  It is still important, however, to firmly establish
the thermal history of the ICM before observed evolution in X-ray luminosity or
temperature can be used to determine cosmological parameters.

\begin{figure}
\begin{center}\leavevmode
\epsfysize=8cm
\epsfbox{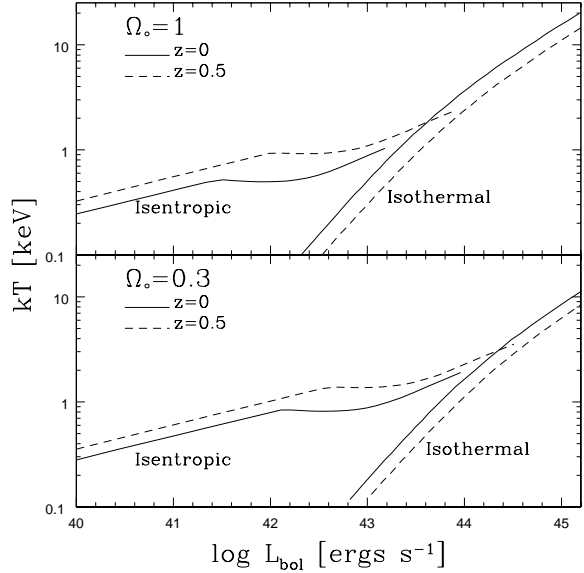}
\end{center}
\caption{The evolution of the luminosity--temperature relation is shown for $\Omega_\circ=1$ (top)
and $\Omega_\circ=0.3$ for both isothermal and isentropic models discussed in the text.
The solid line represents the relation at z=0, and the dashed line the relation at z=0.5.
The (mild) evolution is in opposite directions in the two models.
\label{fig-LTz}}
\end{figure}

The direction of evolution of the L--T relation, shown in Figure \ref{fig-LTz},
depends on the gas physics involved.  In the isentropic model, clusters of a given
temperature are significantly less luminous at z=0.5 than they are at z=0, whereas, 
in the isothermal case,
they are somewhat more luminous.  The work of Henry (1997) shows little or no evolution
in the L--T relation for $3<kT<10$ keV, in confirmation of earlier work 
\cite{HJG,D96,MS97}.  The prediction 
of moderate evolution in the isothermal model
was part of the impetus for considering models with pre-heated gas \cite{K91,EH91};
unfortunately, again, our isentropic model is not valid at these temperatures.  Since errors
in temperature tend to dominate observations, it is unlikely that the predicted evolution
of the L--T relation on group scales will be easily observed.

\begin{figure}
\begin{center}\leavevmode
\epsfysize=8.4cm
\epsfbox{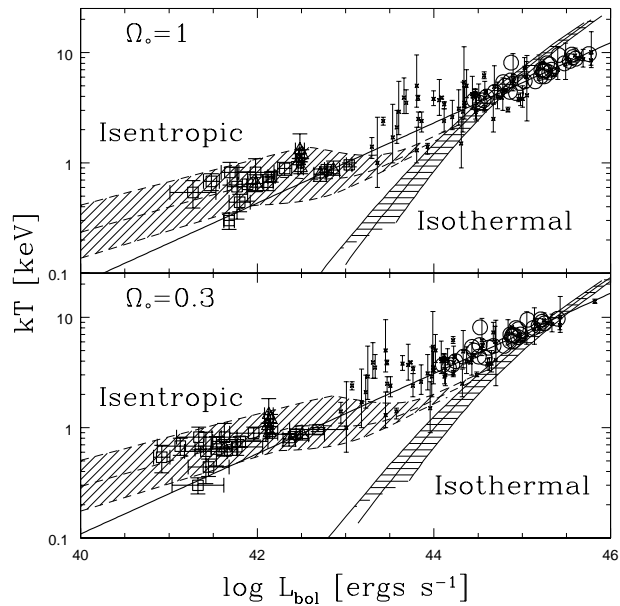}
\end{center}
\caption{This shows the same models and data as presented in Figure \ref{fig-LT}, but assuming
a larger universal baryon fraction of $\Omega_b/\Omega_\circ=0.1$.  For $\Omega_\circ=1$, the
match between the isentropic model and the data is substantially improved.
\label{fig-LT_hib}}
\end{figure}

Many observations have suggested that the contribution of X-ray gas to the total
mass in galaxy clusters is 10--20 per cent,  significantly larger than the Big Bang nucleosynthesis value
assumed here \cite{W+93,WF95,WJF}.  Increasing the baryon density
of the universe allows the application of our isentropic model to more massive haloes, as
the limits in Equations \ref{eqn-Vcrit} and \ref{eqn-Vmax} are increased.    We explore, in
Figure \ref{fig-LT_hib}, the luminosity--temperature relation assuming $\Omega_b/\Omega_\circ=0.1$.
(Below our critical mass,
Equation \ref{eqn-Vcrit}, the gas mass fraction is still given by Equation \ref{eqn-bondi} and, hence, less
than $\Omega_b/\Omega_\circ=0.1$, in qualitative accord with observations \cite{DJF}).
This requires a slight modification of the entropy constant to $K_\circ=0.35K_{34}$ for $\Omega_\circ=1$
and  $K_\circ=0.39K_{34}$ for $\Omega_\circ=0.3$.  The improvement of the fit to the data is
quite noticeable in the $\Omega_\circ=1$ case, and is due primarily to the increase of 
the upper mass limit at which the gas fraction reaches $\Omega_b/\Omega_\circ$.

The isentropic model requires that the infalling gas have an initial entropy given
by $K_\circ=(0.37\pm 0.1)K_{34}$.  This is consistent, with about 1$\sigma$ confidence,
 with the observed value recently
published by Ponman, Cannon, Navarro (1998), $K_\circ=(0.20\pm 0.1)K_{34}$.  To generate this entropy
requires some mechanism for heating the gas (at the turnaround time) from the
background temperature of $10^4$ K to 
about $3 \times 10^6$ K (see \S~\ref{sec-LT}); the 
energy required to heat a total mass of gas $\Omega_b M$ from $\approx 10^4$ K to this
temperature is $1.2 \times 10^{61} (\Omega_b M/10^{13}M_\odot)$ ergs.  If the amount of energy
of a single SN event is about $10^{51}$ ergs, 10 per cent of this is converted into
thermal energy of the hot gas \cite{BR92}, and the mass in stars is about 10 percent of the
mass in gas, \cite{W+93}, this requires a SN rate of $1.2 \times 10^9$ events in the
lifetime of a galaxy with stellar mass $10^{10} M_\odot$.  Over the likely timescales of
galaxy formation, this is probably an unreasonably high rate.

Both models explored here fail to reproduce the temperature function, luminosity function
and L--T relation for the majority of massive clusters.  This mass range is
successfully described in some aspects 
by the model of Evrard \& Henry (1991), in which only the core gas retains its initial
entropy, and the possibility of
mergers stirring up and mixing the gas is ignored.  An alternative model which may be at least
equally successful is one in which the gas is more thoroughly mixed, and behaves like
a polytrope with $\gamma$ varying from 5/3 on group scales to 1.1--1.2 on cluster scales,
and approaching unity as the systems grow increasingly massive. Such a model has been postulated as an explanation of
numerical simulation results (Lewis et al., in preparation), observed
temperature gradients \cite{MFSV} and the discrepancy between X-ray determined masses
and those obtained from lensing observations \cite{MEB}.  We are currently exploring
the predictions of such a model under the present formalism.  

\section{Summary and Conclusions}\label{sec-conc}
Two extreme models of hot intra-cluster and intra-group gas have been considered in
detail.  In the first model, the gas is distributed isothermally at the virial
temperature of the mass.  In the isentropic model, the 
adiabatic collapse of pre-heated gas onto an
isothermal potential is considered, with the gas content of the halo constrained by
the physics of adiabatic Bondi accretion.
In both models we incorporate the Lacey \& Cole (1993; 1994) distribution of halo 
formation times, and also the contribution of recombination radiation to the gas
temperature and luminosity.  The important conclusions are the following:
\begin{itemize}
\item {Neglecting the distribution of halo formation times leads to a 10--20 per cent
overestimate of the normalisation parameter $\sigma_8$}
\item {Neglecting the contribution from recombination radiation has a significant
effect on the slope of the luminosity--temperature relation, for $kT<4$ keV.}
\item {In the isentropic model, we assume the entropy of the pre-infall
gas has been increased by some, unspecified mechanism.  For haloes with X-ray gas temperatures $kT<1$ keV,
the luminosity--temperature relation in this model is $L \propto T^5$.  Both the
slope and dispersion (due to the distribution of formation times) 
of this model are a good match to the available X-ray observations.  
 The fit is improved, especially for $\Omega_\circ=1$,
if we assume a baryon fraction of $\Omega_b/\Omega_\circ=0.1$.}
\item{The isentropic model predicts that the gas temperature in haloes with $M<10^{12}M_\odot$ 
will be significantly higher than the virial temperature; thus, making the assumption that
the gas is at $T_{\rm vir}$ will lead to an overestimation of the mass by more than an order
of magnitude.  This discrepancy can be reduced somewhat if the gas accretion rate is
lower than the critical adiabatic Bondi rate we have assumed here; however, we still predict
that the gas fraction will increase proportionally to the halo mass in this regime.}
\item{The amount of heating required at $z\approx1$ in the isentropic model is equivalent
to heating halo gas at the turnaround epoch to $3 \times 10^6$ K for $\Omega_\circ=1$
and $2 \times 10^6$ K for $\Omega_\circ=0.3$. }
\item{The slope of the observed luminosity function is well matched by the isentropic model for
$L<10^{43}$ ergs s$^{-1}$}
\item {The isentropic model fails for more massive clusters, where the gas entropy
becomes increasingly dominated by shocks.}
\end{itemize}

As better X-ray observations become available for less luminous, cooler groups of galaxies,
the observed slope of the L--T relation in this regime will provide a strong test of the
validity of the isentropic model.  For the majority of clusters, the gas is clearly not
isentropic or isothermal.  These objects may be best described either by largely isothermal
gas with a high entropy core \cite{EH91}, by the shock-heating model of
Cavaliere et al. (1997), or by a model which treats the gas as a uniformly mixed polytrope
with $\gamma \approx 1.2$.

\section*{Acknowledgements}
We are grateful to NYU, where this work was initiated, for their accommodation
and hospitality.  We would like to acknowledge many fruitful discussions with Nick
Kaiser and Julio Navarro, and we are grateful to Ue--Li Pen for providing his unpublished
data.   MLB appreciates critical readings by J. Navarro and the referee, T. Ponman, 
which led to substantial clarifications and improvements.  AB wishes to acknowledge partial support from 
NASA Astrophysics Theory Grant NAG 5-4242.
MLB and DRP are supported in part by Natural Sciences and Engineering 
Research Council of Canada (NSERC)
research grants to C. J. Pritchet and A. Babul.  MLB is also supported by an NSERC 
postgraduate scholarship. 

\appendix
\section{}\label{sec-appendix}
In this appendix we will discuss two different models in which pre-heated gas collapses
adiabatically.  We will make two assumptions about the behaviour of 
$\rho_{\rm g,R}$ and $K_\circ$ that are different from those which we make for
our isentropic model, discussed in \S~\ref{sec-adiabat}.  In both cases, we assume
that $\rho_{\rm g,R} / \rho_R=\Omega_b / \Omega_\circ$, as in the isothermal model.

First, we can assume that  $K_\circ$ is a universal constant, independent of halo mass. 
Then, from Equations \ref{eqn-A} and \ref{eqn-mratios}, 
the ratio of gas to dark matter mass within $R_{\rm vir}$ will
vary strongly with $V_c$, in the sense that more massive clusters will have smaller
gas fractions.  The gas mass fraction in this case would vary by two orders of magnitude
between haloes with $V_c=300$ km s$^{-1}$ and those with $V_c=1000$ km s$^{-1}$.  However, both observations and theory
suggest that the baryon fraction in clusters may be greater than the universal mean,
but do not deviate from this value by more than a factor of a few
\cite{W+93,WF95,DJF,E97}.  Thus, we reject this 
model as improbable on physical grounds.
 
Alternatively, we can constrain the ratio of gas mass to
dark matter mass to be a constant equal to the universal value,
$M_g / M = \Omega_b / \Omega_\circ$.  Then
from Equation \ref{eqn-mratios}, A=1.03 is independent of mass and $z_f$, which
requires the entropy constant 
to scale as $K_\circ \propto V_c^2$ (from Equation \ref{eqn-A}).  The entropy
of the gas will still be constant for any given cluster, but 
will be higher for larger haloes; this requires a mechanism for pre-heating
the gas that is more efficient for more massive haloes.
In this case, since the parameter A is constant, 
$T_{\rm em} \propto T_{g,R} \propto V_c^2 \propto T_{\rm vir}$ from Equations
\ref{eqn-To} and \ref{eqn-ToTem} and, thus, the isothermal mass--temperature relation
($T \propto M^{2/3}$, Equation \ref{eqn-MTIso}) still holds, apart from the normalisation. 
 Furthermore, from Equation 
\ref{eqn-Alum}, recalling that $K_\circ \propto V_c^2$, it holds that 
$L\propto V_c^4$ when bremsstrahlung radiation is the dominant emission
mechanism and $L\propto V_c^2$ when recombination radiation dominates; consequently, $L\propto T^2$ and 
$L\propto T$ for the two cases, respectively.  Interestingly, the above scaling relations are 
exactly the same as those derived in the isothermal case (\S~\ref{sec-isothermal}).  The relationships
are {\em not unique} to the isothermal model.

\end{document}